\documentclass[12pt]{article}

\usepackage{amssymb,amsmath}

\usepackage{graphicx}
\DeclareGraphicsExtensions{.pdf,.png,.jpg}

\usepackage{xcolor} 

\newcommand{\beq}[1]{\begin{equation}\label{#1}}
\newcommand{\eeq}{\end{equation}}
\newcommand{\bear}[1]{\begin{eqnarray}\label{#1}}
\newcommand{\ear}{\end{eqnarray}}

\newcommand{\R}{ {\mathbb R} }

\renewcommand{\theequation}{\arabic{section}.\arabic{equation}}
 \catcode`\@=11 \@addtoreset{equation}{section}\catcode`\@=12

\begin{document}

 \begin{center}
 \large \bf
Generalized Ellis-Bronnikov  wormhole solution in the scalar-Einstein-Gauss-Bonnet $4d$ gravitational model

 \end{center}

 \vspace{0.3truecm}

 \begin{center}

 \normalsize\bf

 \vspace{0.3truecm}

   K. K. Ernazarov\footnote{e-mail: kubantai80@mail.ru (corresponding author)} \footnote{Institute of Gravitation and Cosmology, \\
    Peoples' Friendship University of Russia (RUDN University), \\
    6 Miklukho-Maklaya Street,  Moscow, 117198, Russian Federation }

\vspace{0.3truecm}
  

  \end{center}

\begin{abstract}
    
We consider the sEGB $4d$ gravitational model with a scalar field  $\varphi\left(u\right)$, Einstein and Gauss-Bonnet terms. The model action contains a potential term $U\left(\varphi\right)$, a Gauss-Bonnet coupling function $f\left(\varphi\right)$ and a parameter $\varepsilon = \pm 1$, where $\varepsilon = 1$ corresponds to the usual scalar field, and $\varepsilon = -1$ to the phantom field. In this paper, the sEGB reconstruction procedure considered in our previous paper is applied to the metric of the Ellis-Bronnikov solution, which describes a massive wormhole in the model with a phantom field (and zero potential). For this metric, written in the Buchdal parameterization with a radial variable $u$, we find a solution of the master equation for $ f\left(\varphi\left(u\right)\right)$ with the integration (reconstruction) parameter $C_0$. We also find expressions for  $U\left(\varphi\left(u\right)\right)$ and $\varepsilon \dot{\varphi}^2 = h\left(u\right)$ for $\varepsilon = \pm 1$. We prove that for all non-trivial  values of the  parameter $C_0 \neq 0$ the function $h\left(u\right)$ is not of constant sign for all admissible $u \in \left(-\infty , +\infty\right)$. This means that for a fixed value of the parameter $\varepsilon = \pm 1$  there is no non-trivial sEGB reconstruction in which the scalar field is a purely ordinary field ($\varepsilon = 1$) or a purely phantom field ($\varepsilon = - 1$).

\end{abstract}




\section{Introduction}

The Einstein-Rosen bridge is obtained by solving the vacuum Einstein equations of the general theory of relativity for the static spherically symmetric case \cite{Ein_Rosen}. It represents a non-traversable wormhole, encumbered by event horizons. Decades later, when considering the Einstein equations in the presence of a scalar field, Ellis \cite{Ellis} and Bronnikov \cite{Br} were able to obtain traversable wormhole solutions, provided they employed a non-standard scalar field, i.e., a phantom field. The necessity of the presence of a phantom field or, more generally, some form of exotic matter violating the energy conditions in classical general relativity was nicely discussed by Morris and Thorne  \cite{Morris_1}, who also contemplated the use of wormholes for rapid interstellar travel.

The properties of Lorentzian wormholes in dilatonic Einstein-Gauss-Bonnet theory in four spacetime dimensions are particularly important. These wormholes do not need any form of exotic matter for their existence. A subset of these wormholes is shown to be linearly stable with respect to radial perturbations. Their domain of existence, and derive a generalised Smarr relation for these wormholes are studied in \cite{Kanti_12}. Novel wormholes are obtained in Einstein-scalar-Gauss-Bonnet theory for several coupling functions. The wormholes may feature a single-throat or a double-throat geometry and do not demand any exotic matter. The scalar field may asymptotically vanish or be finite, and it may possess radial excitations. The domain of existence is fully mapped out for various forms of the coupling function \cite{Antonio_12}.

In General Relativity the non-trivial topology of traversable wormholes can be achieved by means of phantom fields, allowing for static and rotating Ellis wormholes, when no further fields are present. In particular, the rotating Ellis wormholes possess many interesting properties, e.g., they satisfy a Smarr relation, they possess as limiting configuration an extremal Kerr black hole, they possess bound orbits, etc. Scalarized wormholes in Scalar-Tensor theory have been considered in Ref. \cite{Xiao_1}. There was shown, that once the corresponding (non-scalarized) wormhole solutions are known in General Relativity, it is no longer necessary to solve the Einstein equations for the metric. But only the equations for the gravitational scalar field and the phantom scalar field need to be solved. Indeed, each solution in General Relativity, and thus each solution for the metric, is characterized by a constant $D$, which relates the charges $Q_{\phi}$ and $Q_{\psi}$ of the gravitational scalar field and the phantom scalar field, respectively, in the Einstein frame, $D^2 = Q^2_{\psi} – Q^2_{\phi}$. Rotating wormhole solutions supported by a complex phantom scalar field with a quartic self-interaction have been done in Ref. \cite{Xiao_2}, where the phantom field induces the rotation of the spacetime. The solutions are regular and asymptotically flat. A subset of solutions described static but not spherically symmetric wormholes is also obtained.

In Ref. \cite{Hyat} investigated the images of asymmetric wormholes as they appear in the different asymptotic regions. Of particular interest here is the presence of a light ring on only one side of the wormhole throat, which can cause light rays to be reflected by the wormhole, resulting in a typical optical/infrared/radio appearance as compared to other compact objects. To demonstrate this phenomenon, there asymmetric Ellis-Bronnikov wormhole as a primary example. The wormhole throat, which connects Universe I and Universe II, is located at $r = - m$, and only Universe I has a light ring. These results show that the behavior of null geodesics in Universe I is similar to that of null geodesics of a Schwarzschild black hole, suggesting that the wormhole appearance on this side can mimic a black hole. However, Universe II has no light ring, and null geodesics starting from infinity in Universe II can pass the wormhole throat into Universe I, and then be reflected back into Universe II.

This paper is a continuation of our research \cite{Ernaz_Ivash} on the reconstruction procedure for the general static spherically symmetric metric written in the Buchdal parameterization of the radial coordinate $u$: $ds^2 = \frac{1}{A\left(u\right)}du^2 - A\left(u\right)dt^2 + C\left(u\right)d{\Omega}^2$, with given $A\left(u\right) > 0$ and $C\left(u\right) > 0$.

Scalar-Einstein-Gauss-Bonnet (sEGB) 4d-gravity is a modified theory of gravity that extends General Relativity (GR) by introducing a coupling between a scalar field and the Gauss-Bonnet (GB) topological invariant. This theory is part of a broader class of scalar-tensor theories and higher-curvature gravitational models, often studied in the context of cosmology, quantum gravity phenomenology, black hole and wormhole physics. Theoretical studies in Scalar-Einstein-Gauss-Bonnet (sEGB) 4d-gravity  are based on very cumbersome mathematics and final solutions often require numerical methods and computer simulations. The "reconstruction method" is based on a second-order differential equation and in some special cases it is possible to find analytical solutions to these equations. By this way, we were able to find exact black hole and wormhole solutions using the "reconstruction method".

Although General Relativity (GR) has been well tested with some astronomical observations, it is known that recent discoveries in astrophysics show that it is not a cornerstone theory. Modifications of General Relativity are mainly motivated by cosmology. There are many unsolved problems in the standard model of cosmology, including: explaining the nature of dark energy and dark matter, creating an accurate model for inflation, or solving the problem of the initial singularity. All these open problems, together with theoretical reasons such as the non-renormalizability of General Relativity, require a more accurate modified gravitational theory. One of the reliable ways of modifications of general relativity (GR) are scalar-tensor theories (ESTT), where the usual Einstein-Hilbert action is supplemented by all possible algebraic second-order curvature invariants with a dynamical scalar field non-minimally coupled to these invariants \cite{Berti}. One way of ESTT is extended scalar-tensor Gauss-Bonnet gravity (ESTGB), for which the scalar field is coupled to a Gauss-Bonnet invariant. The ESTGB gravity field equations are of the second order, as in general relativity, unlike the general ESTT, where even higher orders are possible. Among the ESTGB gravity models, the most studied in the last decade is the Einstein-dilaton-Gauss-Bonnet (EdGB) gravity variant, which is characterized by the coupling function $ \alpha e^{\gamma\varphi}$ for the dilaton field, where $\alpha$ and $\gamma$ are constants. Non-rotating EdGB black holes have been studied perturbatively or numerically in \cite{Mig_2} - \cite{Maeda_4}. It has been shown that EdGB black holes exist when the black hole mass is larger than a certain lower bound proportional to the parameter $\alpha$. Based on the EdGB model, slowly rotating black holes and their properties have been studied. A large number of papers have been published in this area, see \cite{Ayz_5} - \cite{Kunz_10} and the references therein. But for rapidly rotating EdGB black holes, results have so far been obtained using numerical methods with small errors. As these results show, rotating EdGB black holes can exist only when the mass and angular momentum fall within a certain region depending on the coupling constant. One additional fact about EdGB black holes is that they can exceed the Kerr bound for angular momentum. The properties, stability and quasi-normal modes of EdGB black holes based on the model were further studied in numerous articles. Dynamical evolution in Gauss-Bonnet gravity and various aspects of collapse were investigated in \cite{Benk_11} - \cite{Vittorio_16}.

In the context of the ESTGB gravity model, the properties of gravitational wave propagation in cosmological and astrophysical spacetime were studied in the context of Einstein-Gauss-Bonnet gravity \cite{Odin_15}. These studies show that based on the ESTGB gravity model, it is possible to construct a theory that can describe the speed of a gravitational wave equal to the speed of light in vacuum, or at least the speed can be compatible with some constraints. However, the propagation of gravitational waves in a spherically symmetric spacetime violates these constraints, so it is impossible for a gravitational wave propagating in a spherically symmetric spacetime to have a propagation speed equal to the speed of light in vacuum.

\section{The  scalar-Einstein-Gauss-Bonnet model}

The scalar-Einstein-Gauss-Bonnet action in four dimensions is given by \cite{Ernaz_Ivash}, \cite{NN}
\begin{eqnarray}
S = \int d^4z\left|g\right|^{\frac{1}{2}} \Bigg(\frac{R\big(g\big)}{2\kappa^2} - \frac{1}{2} 
 \varepsilon g^{MN}\partial_M\varphi\partial_N\varphi - U\left(\varphi\right)
  + f\left(\varphi\right) {\cal G} \Bigg) , \label{1A} 
\end{eqnarray}
where  $\kappa^2 = 8\pi\frac{G_N}{c^4}$ ($G_N$ is Newton's gravitational constant, $c$ is speed of light), 
$\varphi$ is scalar field, $g_{MN} dz^M dz^N$ is $4d$ metric of signature $(-,+,+,+)$, 
$R\left[g\right]$ is scalar curvature, ${\cal G}$ is Gauss-Bonnet term, $U(\varphi)$ is potential, 
 $f(\varphi)$ is coupling function and $\varepsilon = \pm 1$. For ordinary scalar field we have $ \varepsilon =  1$,  while for phantom one we should put $\varepsilon = - 1$.

We study the spherically-symmetric solutions with the metric
\begin{eqnarray}
ds^2 = g_{MN}(z)dz^Mdz^N = e^{2\gamma\left(u\right)}du^2 - e^{2\alpha\left(u\right)}dt^2
 + e^{2\beta\left(u\right)}d\Omega^2  \label{2A} 
\end{eqnarray}
defined on the manifold
\begin{equation}
   M = \R \times \R_{*} \times S^2.    \label{3A}
\end{equation}
Here $\R_{*} = \left(2\mu, +\infty \right)$ and $S^2$ is $2$-dimensional sphere with the metric 
$d\Omega^2 = d\theta^2 + sin^2\theta d\varphi^2$, where $ 0 < \theta < \pi$ and $0 < \varphi < 2\pi$.

By substitution the metric  (\ref{2A}) into the action we obtain 
\begin{equation}
 S = 4 \pi \int du \left(L + \frac{dF_{*}}{du} \right),  \label{4AA}   
\end{equation}
where the Lagrangian $L$ reads
\begin{eqnarray}
L = \frac{1}{\kappa^2}\Bigg(e^{\alpha- \gamma + 2\beta}\dot{\beta}\left(\dot{\beta}
 + 2\dot{\alpha}\right) + e^{\alpha + \gamma}\Bigg) \nonumber \\
  - \frac{1}{2}e^{\alpha - \gamma + 2\beta}  \varepsilon \dot{\varphi}^2
   - e^{\alpha + \gamma + 2\beta}U\left(\varphi\right)
    - 8\dot\alpha\dot\varphi\frac{df}{d\varphi}\Bigg(\dot{\beta}^2e^{\alpha + 2\beta
    - 3\gamma} - e^{\alpha - \gamma}\Bigg), \label{4A} 
\end{eqnarray}
and the total derivative term is irrelevant for our consideration.
Here and in what follows we denote $\dot{x} = \frac{dx}{du}$. 

By using (without loss of generality) the Buchdal radial gauge obeying $\alpha = - \gamma$. 
For the metric (\ref{2A}) we obtain
\color{black}
\begin{equation}
ds^2 = \left(A\left(u\right)\right)^{-1}du^2 - A\left(u\right)dt^2 + C\left(u\right)d\Omega^2,  
\label{5Buch} 
\end{equation}
where
\begin{equation}
e^{2\gamma\left(u\right)} = \left(A\left(u\right)\right)^{-1}, \quad
e^{2\alpha\left(u\right)} = A\left(u\right) > 0,\quad
e^{2\beta\left(u\right)} = C\left(u\right) > 0.
 \label{5AC}
\end{equation}

In what follows we use the identities
\begin{equation}
\dot{\alpha}= \frac{\dot{A}}{2A}, \qquad  \dot{\beta}= \frac{\dot{C}}{2C}.  \label{5AB}
\end{equation}

 We put (without loss of generality) $\kappa^2 = 1$. We also denote
\begin{equation}
   f \left(\varphi\left(u\right)\right) = f, 
  \qquad  U \left(\varphi\left(u\right)\right) = U  \label{6_FU}
\end{equation}
and hence
\begin{equation}
\frac{d}{du}f = \frac{df}{d\varphi}\frac{d\varphi}{du} \Longleftrightarrow \dot{f}
 = \frac{df}{d\varphi}\dot{\varphi},  \label{6_F} 
\end{equation}

\begin{equation}
\frac{d}{du}U = \frac{dU}{d\varphi}\frac{d\varphi}{du} \Longleftrightarrow \dot{U}
 = \frac{dU}{d\varphi}\dot{\varphi}.  \label{6_U} 
\end{equation}

A more detailed solution of the Lagrange equation (\ref{4A}) and an extensive analysis of the solutions were made in our previous article \cite{Ernaz_Ivash}. From the Lagrangian (\ref{4A})  we get following set of equations.  
\color{black}
\begin{equation}
\dot{A}\left[8\dot{f}\left(1 - 3KA\right) + \dot{C} \right]
 + 2KA - 2 - CA  \varepsilon \dot{\varphi}^2 + 2CU = 0,  \label{6G} 
\end{equation}
where here and in what follows we use the notation
\begin{equation}
K \equiv \left(\frac{\dot{C}}{2C}\right)^2C.  \label{6K} 
\end{equation}

\begin{eqnarray}
\begin{gathered}
16\ddot{f}A\left(1 - KA\right) + 8\dot{f}\left(\dot{A} - 3KA\dot{A} - 2\dot{K}A^2\right) + \dot{A}\dot{C} \\
 + 2A\left(\ddot{C} - K \right) + CA
  \varepsilon \dot{\varphi}^2 - 2 + 2CU = 0.  \label{6A} 
\end{gathered}
\end{eqnarray}

\begin{eqnarray}
\begin{gathered}
\left(C - 4\dot{f}A\dot{C}\right)\ddot{A} - 4\ddot{f}\dot{A}A\dot{C}
 - 4\dot{f}\Bigg(\left(\dot{A}\right)^2 \dot{C} + \dot{A}A\ddot{C} - 2\dot{A}AK\Bigg)  + \\
 + \dot{A}\dot{C} 
 + C\left(A  \varepsilon \dot{\varphi}^2 + 2U\right) + A\ddot{C} - 2AK = 0.  \label{6B} 
\end{gathered}
\end{eqnarray}

\begin{eqnarray}
4\dot{f}\left(AK - 1\right)\ddot{A} + 
 \varepsilon \ddot{\varphi} \dot{\varphi}AC
  + 4\dot{f}\dot{A}\left(\dot{A}K + A\dot{K}\right) 
\nonumber \\
+ \left(\dot{A}C + A\dot{C}\right)
 \varepsilon \dot{\varphi}^2 - C \dot{U} = 0.  \label{6P} 
\end{eqnarray}

Here we put the following restriction
\begin{eqnarray}
\begin{gathered}
\dot{\varphi} \neq 0 \quad $ for$ \quad u \in \left(u_{-}, u_{+}\right), \label{7PN} 
\end{gathered}
\end{eqnarray}
where  interval $\left(u_{-}, u_{+}\right)$ is belonging to $\R_{*} = \left(2\mu, +\infty \right)$. 

By adding equations (\ref{6A}) and (\ref{6G}) and dividing the result by $4C$  we get the relation for the 
 function $U = U\left(\varphi(u)\right)$
\begin{eqnarray}
\begin{gathered}
U = \frac{1}{C}\Bigg(1 - 4A\left(1 - KA\right)\ddot{f}
 - \dot{A}\left[4\dot{f}\left(1 - 3KA\right) + \frac{1}{2}\dot{C}\right] + 4\dot{f}\dot{K}A^2 - \frac{1}{2}A\ddot{C}\Bigg).  \label{6U} 
\end{gathered}
\end{eqnarray}

The relation (\ref{6U}) may be written as
\begin{eqnarray}
\begin{gathered}
C U = E_{U}\ddot{f} + F_{U}\dot{f} + G_U , \label{6UU} 
\end{gathered}
\end{eqnarray}
where
\begin{eqnarray}
E_U = -4A\left(1 - KA\right),  \label{6EU} \\
F_U = -4\dot{A}\left(1 - 3KA\right) + 4\dot{K}A^2, \label{6FU} \\
G_U = 1 - \frac{1}{2}\dot{A}\dot{C} - \frac{1}{2}A\ddot{C}. \label{6GU} 
\end{eqnarray}
 
Subtracting (\ref{6G}) from (\ref{6A}) and dividing the result by $2A$, 
we obtain the relation for $\dot{\varphi}$
\begin{eqnarray}
\begin{gathered}
C  \varepsilon \dot{\varphi}^2
 = 8\ddot{f}\left(KA - 1 \right) + 8\dot{f}\dot{K}A + 2K - \ddot{C}
\equiv \Phi_{\varepsilon}. \label{6P2}
 \end{gathered}
\end{eqnarray}
agreement with the relation $(14)$ from Ref. \cite{NN}. 
 
 Due to  $C\left(u\right) > 0$ and (\ref{7PN}) we get 
\begin{eqnarray}
\begin{gathered}
\varepsilon \Phi_{\varepsilon}  > 0 \label{6Phi} 
\end{gathered}
\end{eqnarray}
for all $u \in \left(u_{-}, u_{+}\right)$.
 
 In the simplest case $C\left(u\right) = u^2$ and $ \varepsilon = 1$  explored in Ref. \cite{NN} we get $K = 1$, $\ddot{C} = 2$ and hence the restriction   (\ref{6Phi}) reads
\begin{eqnarray}
\begin{gathered}
 8\ddot{f}\left(A - 1\right) = \Phi_{1}/u^2 > 0.  \label{7Phi} 
\end{gathered}
\end{eqnarray}

Subtracting (\ref{6A}) from (\ref{6B}), we get the master equation for the  
function $f = f(\varphi(u))$
 \begin{equation}
  E\ddot{f} + F\dot{f} + G = 0,  \label{6FF} 
 \end{equation}
where 
 \begin{eqnarray}
 E = 4A\left(4KA - \dot{A}\dot{C} - 4\right),  \label{6E} \\
F = -4\ddot{A}A\dot{C} - 4\left(\dot{A}\right)^2\dot{C}
 - 4\dot{A}A\ddot{C} + 8\left(4KA\dot{A} 
 - \dot{A} + 2\dot{K}A^2\right), \label{6F6} \\
 G = C\ddot{A} - A\ddot{C} + 2. \label{6G6} 
\end{eqnarray}

In Ref.  \cite{NN} the main results were obtained for the specific case, when $\varepsilon = 1$  and $C\left(u\right) = u^2$ and these results completely coincide with above results.

\color{black}

\section{Solutions to master eqution  }

Here we consider the solutions to nonhomogenous linear equation of the second order (\ref{6FF}).  Let us consider the general case, when 
\begin{equation}
G\left(u\right) \neq 0 \quad {\rm for} \quad u \in \left(u_{-}, u_{+}\right), \label{7E} 
\end{equation}
where $\left(u_{-}, u_{+}\right)$ is interval from (\ref{7PN}). 

On introducing a new function $z = \dot{f}$ into (\ref{6FF}), we get 

\begin{equation}
\dot{z} + \eta\left(u\right)z + \xi\left(u\right) = 0, \label{7.Y} 
\end{equation}
where
\begin{eqnarray}
\begin{gathered}
\eta\left(u\right) = \frac{F\left(u\right)}{E\left(u\right)}, 
\qquad   \xi\left(u\right) = \frac{G\left(u\right)}{E\left(u\right)}. \label{7.AB} 
\end{gathered}
\end{eqnarray}

The solution to differential equation (\ref{7.Y})  is as follows:
\begin{eqnarray}
\begin{gathered}
\dot{f} = z = C_0 z_0\left(u\right) - 
  z_0\left(u\right)\int\limits_{u_0}^u dw \xi\left(w\right)\left(z_0\left(w\right)\right)^{-1}, \label{8.Y} 
\end{gathered}
\end{eqnarray}
where $u \in \left(u_{-}, u_{+}\right)$, $C_0$ is an arbitrary constant and
\begin{eqnarray}
\begin{gathered}
z_0\left(u\right) = \exp\left( - \int\limits_{u_0}^u d v \eta\left(v \right)\right) \label{8.Y0} 
\end{gathered}
\end{eqnarray}
is the solution to homogeneous linear differential equation: $\dot{z_0} + \eta\left(u\right)z_0 = 0$.

Integrating (\ref{8.Y}) we obtain a general solution to (\ref{6FF}) 
\begin{eqnarray}
\begin{gathered}
f =  C_1 + C_0 \int \limits_{u_0}^u dv z_0\left(v\right)
 - \int\limits_{u_0}^u d v z_0 \left(v\right)\int\limits_{u_0}^v dw\xi\left(w\right)\left(z_0\left(w\right)\right)^{-1}, \label{8.F} 
\end{gathered}
\end{eqnarray}
where $C_1$ is an arbitrary constant. 
We note that relation (\ref{6Phi}) impose restrictions only on $C_0$ and $u_0$ since the function 
 $\Phi_{\varepsilon}$ depends on $\dot{f}$ and $\ddot{f}$.


Now let us consider the specific case, when  
\begin{eqnarray}
\begin{gathered}
G\left(u\right) = 0 \quad $for all$ \quad u \in \left(u_{-}, u_{+}\right). \label{7E0} 
\end{gathered}
\end{eqnarray}

In this case the  master equation (\ref{6FF}) has following form
\begin{equation}
 E\ddot{f} + F\dot{f}  = 0  \label{6FFE0} 
 \end{equation}
and in the subcase 
\begin{eqnarray}
\begin{gathered}
E\left(u\right) \neq  0, \qquad F\left(u\right) \neq  0 \quad $for all$ \quad u \in \left(u_{-}, u_{+}\right) \label{7E0} 
\end{gathered}
\end{eqnarray}
the solution to master equation (\ref{6FFE0}) has the following form 
\begin{eqnarray}
\begin{gathered}
f =  C_1 -  C_0 \int \limits_{u_0}^u du z_0\left(u\right), 
\qquad 
\end{gathered}
\end{eqnarray}
where $C_1$ is an arbitrary constant. 

For the subcase, when
\begin{eqnarray}
\begin{gathered}
F\left(u\right) =  E \left(u\right) = 0 \quad $for all$ \quad u \in \left(u_{-}, u_{+}\right), \label{7EF0} 
\end{gathered}
\end{eqnarray}
the master equation has a solution only if 
\begin{eqnarray}
\begin{gathered}
G\left(u\right) =  0 \quad $for all$ \quad u \in \left(u_{-}, u_{+}\right). \label{7G0} 
\end{gathered}
\end{eqnarray}
In this case  function $f$ can be arbitrary one.

\section{Generalized Ellis-Bronnikov wormhole solution}

Here we consider an example of reconstruction procedure to test the reconstruction scheme  under consideration. We apply the reconstruction procedure \cite{Ernaz_Ivash} to the metric of the Ellis-Bronnikov wormhole solution \cite{Sh_Ish}

\begin{equation}
ds^2 = \left(A\left(u\right)\right)^{-1}du^2 - A\left(u\right)dt^2 + C\left(u\right)d\Omega^2,  
\label{EL_Bron_1} 
\end{equation}
where
\begin{eqnarray}
A\left(u\right) = e^{2\alpha\left(u\right)},  \nonumber 
\\  C\left(u\right) = e^{-2\alpha\left(u\right)}\left(u^2 + L^2\right), 
\\ \alpha\left(u\right) = \frac{\mu}{L}\Big({\rm arctan}\frac{u}{L} - \frac{\pi}{2}\Big). \nonumber
\label{EL_Bron_2} 
\end{eqnarray}

Here and below we assume $L>0$ and  $\mu>0$.
Taking into account the following asymptotic behavior:

\begin{eqnarray}
e^{2\alpha\left(u\right)}|_{u \to +\infty} = \Bigg(1 - \frac{2\mu}{u} \Bigg) + O\big(u^{-2}\big), \nonumber \\ 
e^{2\alpha\left(u\right)}|_{u \to -\infty} = e^{-\frac{2\pi\mu}{L}}\Bigg(1 + \frac{2\mu}{\left|u\right|} \Bigg) + O\big(\left|u\right|^{-2}\big), \nonumber
\end{eqnarray}
it can be seen that the space-time with metric (\ref{EL_Bron_1})  has two asymptotically flat domains $R_\pm$ as $ u \to \pm\infty$. In the domain $R_{+}$  the asymptotic mass is defined as $\mu_+=\mu$. Then in the domain $R_-$ the asymptotic mass is equal to

\begin{eqnarray}
\mu_- = -\mu\cdot e^{-\frac{\pi \mu}{L}}. \nonumber
\end{eqnarray}

It is obviously that the masses have both different values and different signs. Therefore, the Ellis-Bronnikov wormhole looks like an object with negative mass for a distant observer in the domain $R_-$. The radius of the throat that connects the asymptotic domains $R_+$ and $R_-$ corresponds to the minimum radius of the two-dimensional sphere

\begin{eqnarray}
R^2\left(u\right) = \left(u^2 + L^2\right) \cdot e^{-2\alpha\left(u\right)}. \nonumber
\end{eqnarray}

At $u_{th} = M$, the radius of the throat reaches its minimum value

\begin{eqnarray}
R_0 = \left(u^2 + L^2\right)^{\frac{1}{2}} \cdot exp\left[-\frac{\mu}{L}\Bigg(arctan\frac{\mu}{L} - \frac{\pi}{2}\Bigg)\right]. \nonumber
\end{eqnarray}

We note that in the case $\mu =0$  we have $\mu_{\pm} = 0$, and the metric  (\ref{EL_Bron_1}) is as follows to the Ellis metric

\begin{equation}
ds^2 = du^2 - dt^2 +  \left(u^2 + L^2\right)d\Omega^2,  
\label{EL_Bron_3} 
\end{equation}

This is a special case of the Ellis-Bronnikov wormhole solution, the reconstruction of which was studied in detail in our previous paper \cite{Ernaz_Ivash}.

In this case, for the master equations (\ref{6FF}) of the functions E(u), F(u) and G(u), defined in (\ref{6E}), (\ref{6F6}) and (\ref{6G6}), are expressed, respectively, as

\begin{equation}
E = 16\Bigg(\frac{\left(u - 2\mu\right)\left(u - \mu\right)}{u^2 + L^2} - 1\Bigg)\cdot exp\Bigg(-\frac{\mu}{L}\bigg(\pi - 2arctan\bigg(\frac{u}{L}\bigg)\bigg)\Bigg),  
\label{EL_Bron_4} 
\end{equation}

\begin{equation}
F = \bigg(\frac{8}{u^2 + L^2}\bigg)^2\bigg(\frac{u}{2} - \mu\bigg)\bigg(\left(u^2 + L^2\right) - \left(u - \mu\right)^2\bigg)\cdot exp\Bigg(-\frac{\mu}{L}\bigg(\pi - 2arctan\bigg(\frac{u}{L}\bigg)\bigg)\Bigg),  
\label{EL_Bron_5} 
\end{equation}

\begin{equation}
G = 0.  
\label{EL_Bron_6} 
\end{equation}

Solving master equation $ E\ddot{f} + F\dot{f} + G = 0$ we obtain

\begin{equation}
f\left(\varphi \left( u \right)\right) = \int\frac{C_0\left(u^2 + L^2\right)}{\bigg(\left(u - \mu\right)\left(u - 2\mu\right) - \left(u^2 + L^2\right)\bigg)^{\frac{2}{3}}} \cdot exp\Bigg(-\frac{2\mu}{L}arctan\bigg(\frac{u}{L}\bigg)\Bigg)du + C_1
\label{EL_Bron_7} 
\end{equation}

and

\begin{eqnarray}
\dot{f} = \frac{C_0\left(u^2 + L^2\right)}{\bigg(\left(u - \mu\right)\left(u - 2\mu\right) - \left(u^2 + L^2\right)\bigg)^{\frac{2}{3}}} \cdot exp\Bigg(-\frac{2\mu}{L}arctan\bigg(\frac{u}{L}\bigg)\Bigg), \nonumber \\
\ddot{f} = \frac{2C_0\left(u - 2 \mu\right)\bigg(\left(u - 2\mu\right)^2 - \left(u^2 + L^2\right)\bigg)}{\bigg(\left(u - \mu\right)\left(u - 2\mu\right) - \left(u^2 + L^2\right)\bigg)^{\frac{5}{3}}} \cdot exp\Bigg(-\frac{2\mu}{L}arctan\bigg(\frac{u}{L}\bigg)\Bigg). 
\label{EL_Bron_8}
\end{eqnarray}

Here $C_0$ and  $C_1$ are arbitrary constants.

In this case, for the scalar potential $U\left(u\right)$ from expressions (\ref{6EU}), (\ref{6FU}), (\ref{6GU}) we obtain the following relationships:

\begin{eqnarray}
E_U = 4\bigg(\frac{ \left(u^2 - \mu \right)^2}{u^2 + L^2} - 1 \bigg) \cdot  exp\Bigg(-\frac{\mu}{L}\bigg(\pi - 2arctan\bigg(\frac{u}{L}\bigg)\bigg)\Bigg), \quad  \label{EL_Bron_9} \\
F_U =- \bigg(\frac{4}{u^2 + L^2}\bigg)^2\bigg(\frac{u}{2} - \mu\bigg)\left(\left(u - \mu \right)^2 - \left(u^2 + L^2\right)\right) \cdot exp\Bigg(-\frac{\mu}{L}\bigg(\pi - 2arctan\bigg(\frac{u}{L}\bigg)\bigg)\Bigg) ,  \quad \label{EL_Bron_10} \\
G_U = 0  \qquad  \label{EL_Bron_11} 
\end{eqnarray}
and hence using (\ref{6U}) and (\ref{EL_Bron_8}) we get the following relation for potential function

\begin{eqnarray}
U =\frac{2C_0\mu\left(u - 2 \mu\right)\left(u - \mu\right)\bigg(\left(u - \mu\right)^2 - \left(u^2 + L^2\right)\bigg)}{\bigg(\left(u - \mu\right)\left(u - 2\mu\right) - \left(u^2 + L^2\right)\bigg)^{\frac{5}{3}}}\bigg(\frac{2}{(u^2 + L^2}\bigg)^2 \cdot exp\Bigg(\frac{2\mu}{L}\left(arctan\bigg(\frac{u}{L}\bigg) - \pi\right)\Bigg). \quad 
\label{EL_Bron_12}
\end{eqnarray}

As $u \to \pm \infty$, the value of the potential function tends to zero. From relation (\ref{6P2}) we get

\begin{eqnarray}
\varepsilon \dot{\varphi}^2 = h\left(u\right),
\label{EL_Bron_13}
\end{eqnarray}

\begin{eqnarray}
h\left(u\right) = -\frac{2}{\left(u^2 + L^2\right)^2}\Bigg(\left(\mu^2 + L^2\right) + \\ \nonumber  \frac{8C_0\mu}{e^{\frac{\pi\mu}{L}}}\frac{\left(\left(6\mu L^2 - 14\mu^3\right)u - \left(L^2 - 15\mu^2\right)u^2 - 4\mu u^3 + 4\mu^4 - 3\mu^2L^2 + L^4\right)}{\left(2\mu^2 - 3\mu u - L^2\right)^{\frac{5}{3}}}\Bigg). \quad  
\label{EB12}
\end{eqnarray}

Let $C_0 \neq 0$  (for $C_0=0$ we have obtained the Ellis-Bronnikov wormhole solution in the scalar-tensor theory without the Gauss-Bonnet term \cite{Ellis, Br} ). From this  formula we obtain asymptotic relations as $u \to \pm\infty$
 
\begin{eqnarray}
h\left(u\right) \sim - \Bigg(\frac{64C_0\mu^{\frac{1}{3}}}{3^{\frac{5}{3}e^{\frac{\pi\mu}{L}}}}\Bigg)\cdot u^{-\frac{8}{3}}.
\label{EL_Bron_15}
\end{eqnarray}

From (\ref{EL_Bron_15}) we see, that for big enough $ \left|u\right|$ the

\begin{eqnarray}
sign\left(h\left(u\right)\right) = - sign\left(C_0\right),
\label{EL_Bron_16}
\end{eqnarray}
which can be written as
\begin{eqnarray}
h\left(u\right) < 0
\label{EL_Bron_17}
\end{eqnarray}
for $C_0 >0$ and
\begin{eqnarray}
h\left(u\right) > 0
\label{EL_Bron_18}
\end{eqnarray}
for $C_0 <0$.

This means that in the case of $C_0 > 0$, due to (\ref{EL_Bron_17}), we obtain a ghost field $\left(\varepsilon = -1\right)$ at a large distance from the neck for big enough $ \left|u\right|$, and in the case of $C_0 < 0$, due to (\ref{EL_Bron_18}), we obtain an ordinary scalar field for big enough $ \left|u\right|$.

At the point

\begin{eqnarray}
u_{\ast} = \frac{2\mu^2 - L^2}{3\mu}
\label{EL_Bron_19}
\end{eqnarray}
the function $h\left(u\right)$ is not formally defined. We calculate the value of the numerator in the second term in brackets on the right-hand side of relation (\ref{EB12}) at the point $u = u_{\ast}$.  In formula (\ref{EB12}) and everywhere in this article we assume

\begin{eqnarray}
x^{\frac{5}{3}} \equiv \left(\sqrt[3]x\right)^5
\label{EL_Bron_20}
\end{eqnarray}
for any real number $x$. Let us denote
\begin{eqnarray}
N\left(u\right) = \left(6\mu L^2 - 14\mu^3\right)u - \left(L^2 - 15\mu^2\right)u^2 - 4\mu u^3 + 4\mu^4 - 3\mu^2L^2 + L^4 
\label{EL_Bron_21}
\end{eqnarray}
and find the value of the function $N\left(u\right)$ at the point $u_{\ast}$. We obtain 

\begin{eqnarray}
N\left(u_{\ast}\right) = \frac{\left(\mu^2 + L^2\right)\left(4\mu^2 + L^2\right)}{27\mu^2} > 0.
\label{EL_Bron_22}
\end{eqnarray}

From this formula follows the following asymptotic relationship for $u \to u_{\ast}$:

\begin{eqnarray}
h\left(u\right) \sim - \frac{1}{\left(u_{\ast}^2 + L^2\right)^2}\cdot \frac{16C_0\mu}{e^{\frac{\pi\mu}{L}}} \cdot \frac{N\left(u_{\ast}\right)}{\left(u_{\ast} - u\right)^{\frac{5}{3}}}
\label{EL_Bron_23}
\end{eqnarray}

Due to (\ref{EL_Bron_20}), the function $\left(u_{\ast} - u\right)^{-\frac{5}{3}} $ tends to $+\infty$  as $u \to u_{\ast} - 0$  (on the left) and to  $-\infty$ as $u \to u_{\ast} + 0$  (on the right). Therefore, from relation (\ref{EL_Bron_23}) we obtain

\begin{eqnarray}
h\left(u\right) \to \left(-sign\left(C_0\right)\right)\infty
\label{EL_Bron_24}
\end{eqnarray}
as $u \to u_{\ast} - 0$  and
\begin{eqnarray}
h\left(u\right) \to \left(sign\left(C_0\right)\right)\infty
\label{EL_Bron_25}
\end{eqnarray}
as $u \to u_{\ast} + 0$.

Thus $u_{\ast}$ is a breaking point of the function $h\left(u\right)$. From these considerations the following statement is followed.

\textbf{Theorem (no-go).} \textsl{There is no such reconstruction constant $C_0 \neq 0$   for the Ellis-Bronnikov metric with parameters $L>0$ and  $\mu >0$ that for all values  $u \in \left(-\infty,+\infty\right)$  except the point $u_{\ast}$ the function $h\left(u\right)$ is of constant sign, i.e. $h\left(u\right) >0$ $\left(\varepsilon = 1 \right)$ or $h\left(u\right) <0$ $\left(\varepsilon =-1 \right)$.}


From a physical point of view, this means that for any non-trivial reconstruction $\left(C_0 \neq 0 \right)$ \cite{Ernaz_Ivash} of the Ellis-Bronnikov metric with parameters $L>0$ and $\mu>0$, the scalar field defined in the intervals $\left(-\infty, u_{\ast}\right)$ and $\left(u_{\ast},  +\infty\right)$  cannot be purely ghost or purely normal (non-ghost).

In the case of a trivial reconstruction of the Ellis-Bronnikov metric (with $L>0$ and $\mu >0$) with the reconstruction constant $C_0=0$ the Gauss-Bonnet term gives a zero contribution to the equations of motion and we arrive at the Ellis-Bronnikov wormhole solution with a purely ghost scalar field $\left(\varepsilon =-1 \right)$ and zero potential $U=0$.

It should be noted that a non-trivial reconstruction of the Ellis metric (\ref{EL_Bron_3}) ($L>0$ and $\mu=0$) was considered earlier in \cite{Ernaz_Ivash}.

Figures 1 and 2 show examples for the function $h\left(u\right)$ with $C_0<0$ or $C_0>0$, respectively. 

\begin{figure}[h]
\center{\includegraphics[scale=0.2]{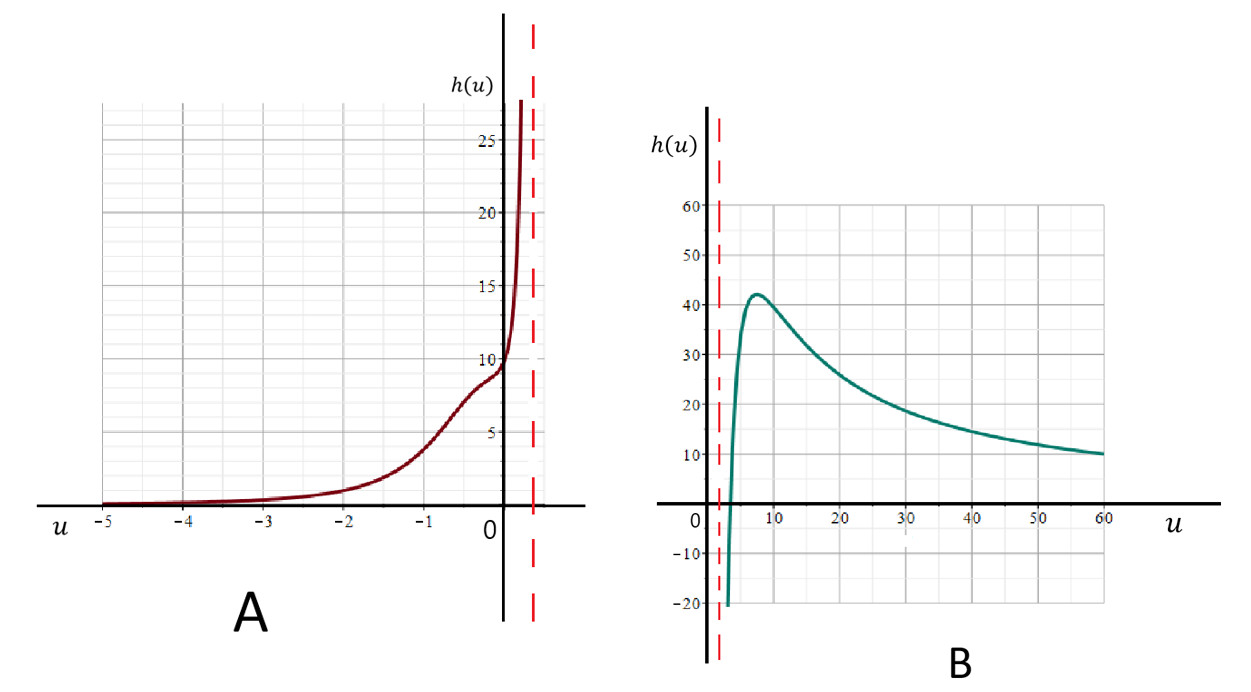}}
\caption{\textsl{The function $h\left(u\right)$   for $\mu = L = 1$ and $C_0 = -10$. A vertical red dashed line crosses point $u_{\ast}=\frac{1}{3}$. A) The function $h\left(u\right)$ is positive in the interval $\left(-\infty, u_{\ast}=\frac{1}{3}\right)$. This means that in this interval a scalar field is ordinary one. B) The function $h\left(u\right)$ is negative in the interval  $\left(u_{\ast}=\frac{1}{3}, u_1=3.4686\right)$ and positive in the interval $\left(u_1=3.4686, +\infty\right)$. Therefore in the interval $\left(u_{\ast}=\frac{1}{3}, u_1=3.4686\right)$ a scalar field is ghost one and in the interval $\left(u_1=3.4686, +\infty\right)$ we obtain a solution with an ordinary field.}} 
\label{Fig_1AB}
\end{figure}

\begin{figure}[h]
\center{\includegraphics[scale=0.2]{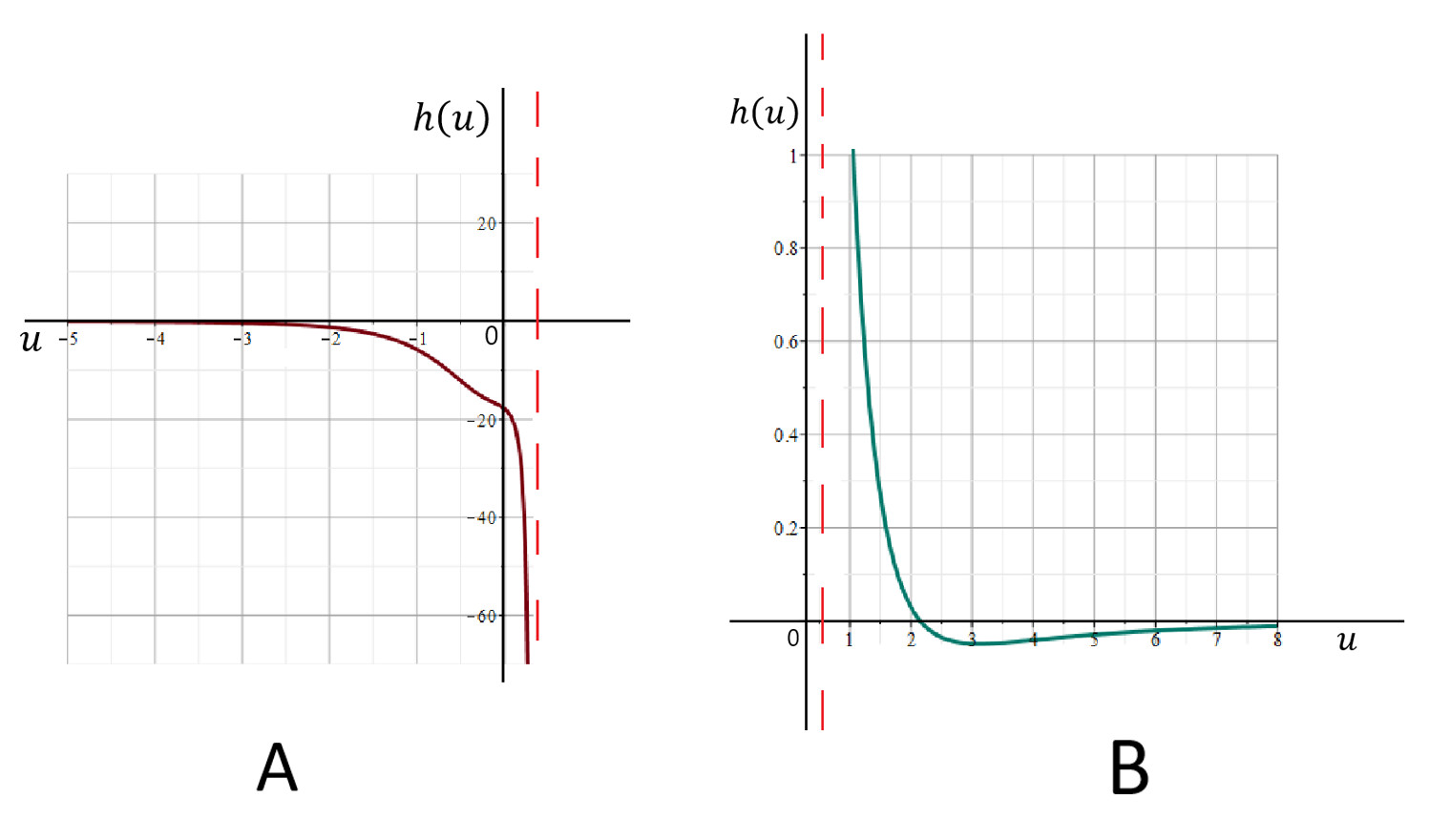}}
\caption{\textsl{The function $h\left(u\right)$  for $\mu = L = 1$ and $C_0 = 10$.  A vertical red dashed line crosses point $u_{\ast}=\frac{1}{3}$. A) The function $h\left(u\right) < 0$ in the interval $\left(-\infty, u_{\ast}=\frac{1}{3}\right)$. This means that in this interval a scalar field is ghost one. B) The function $h\left(u\right)$ is positive in the interval  $\left(u_{\ast}=\frac{1}{3}, u_1=2.1496\right)$ and negative in the interval  $\left(u_1=2.1496, +\infty\right)$. Therefore only in the interval $\left(u_{\ast}=\frac{1}{3}, u_1=2.1496\right)$ we obtain a solution with an ordinary field.}} 
\label{Fig_2AB}
\end{figure}

From Fig.1 we see that, for $\mu=L=1$ and $C_0=-10$  in the interval $\left(u_{\ast}=\frac{1}{3}, u_1=3.4686\right)$ we obtain a solution with a ghost field, and in the intervals $\left(-\infty, u_{\ast}=\frac{1}{3}\right)$ and $\left(u_1=3.4686, +\infty\right)$ we obtain a solution with an ordinary field.

From Fig.2 we see that, for $\mu =L=1$ and $C_0=10$  in the interval $\left(u_{\ast}=\frac{1}{3}, u_1=2.1496\right)$ we obtain a solution with an ordinary field, and in the intervals $\left(-\infty, u_{\ast}=\frac{1}{3}\right)$ and $\left(u_1=2.1496, +\infty\right)$ we obtain a solution with a ghost field.

\section{Conclusions}

In this paper we consider the sEGB model - a $4d$ gravitational model with a scalar field $\varphi$, Einstein and Gauss-Bonnet terms. The model action contains a potential term $U\left(\varphi\right))$, a Gauss-Bonnet coupling function $f\left(\varphi\right)$ and a parameter $\varepsilon = \pm 1$, where $\varepsilon = 1$ corresponds to a pure scalar field, and $\varepsilon = -1$ to a phantom one.

Here we applied the sEGB reconstruction procedure considered in our previous paper to the metric of the Ellis-Bronnikov solution, which describes a massive wormhole in a model with a phantom field (and zero potential). As was pointed out in Ref. \cite{Sharma_14} Ellis-Bronnikov wormholes require violation of null energy conditions at the "throat".

For this metric, written in the Buchdal parameterization with a radial variable $u$, we found a solution of the master equation for $f\left(\varphi\right))$ with the integration (reconstruction) parameter $C_0$. Also, expressions for $U\left(\varphi\left(u\right)\right)$ and $\varepsilon \dot{\varphi} = h\left(u\right)$ were found for $\varepsilon = \pm 1$.
We found that for big enough values of  $\left|u\right|$: the scalar field is an ordinary field in case a) $C_0 < 0$, while it is a phantom one when b) $C_0 > 0$. It should be noted that case a) is more acceptable from a physical point of view. In this case, the value of the field potential $U\left(\varphi\left(u\right)\right)$ tends to zero as $u \to \pm \infty$, independently from the sign of an arbitrary constant $C_0 \neq 0$. Based on the obtained results, a theorem (no-go) was established in this metric, which reported that  does not exist as a constant-sign function $h\left(u\right)$ for all values of $u \in \left(- \infty, + \infty \right)$.

We have proved that for all non-trivial values of the parameter $C_0 \neq 0$ the function $h\left(u \right)$ is not of constant sign for all admissible $u \in \left(- \infty, + \infty \right)$ (no-go theorem). This means that for a fixed value of the parameter $\varepsilon = \pm 1$ there is no non-trivial sEGB reconstruction in which the scalar field is a purely ordinary field ($\varepsilon = 1$) or a purely phantom one ($\varepsilon = -1$).
It should be noted that in our previous paper \cite{Ernaz_Ivash} the sEGB reconstruction procedure was applied to the Ellis wormhole metric, which is a special case of the Ellis-Bronnikov wormhole metric in the case of zero mass parameter $\mu$. We have found that in this case the sEGB reconstruction procedure yields: $\varepsilon = -1$, $U\left(\varphi\right) = 0$ and $ f\left(\varphi\right) = c_1+c_0 \left(tan\left(\varphi\right)+1/3\left(tan\left(\varphi\right)\right)^3\right) $, where $c_1$ and $c_0$ are arbitrary constants. This result is consistent with the result obtained in this paper for $\mu = 0$.

"The reconstruction method" is constructed for the spherical symmetric metric (for non-rotating black hole and wormhole models) taking into account the Buchdal parameterization. For rotating black holes and spinning wormholes, another, more hard reconstruction method is required.

It is of interest to generalize the results of this work to the multidimensional case.

\section*{Acknowledgements}

I thank V.D. Ivashchuk for helpful discussions and comments on the draft. The research was funded by RUDN University, scientific project number FSSF-2023-0003.    

\section{Data Availability Statements}

No Data associated in the manuscript.

\renewcommand{\theequation}{\Alph{subsection}.\arabic{equation}}
\renewcommand{\thesection}{}
\renewcommand{\thesubsection}{\Alph{subsection}}
\setcounter{section}{0}

\section{Appendix}

\subsection{Energy conditions}

\indent To analyze the energy condutions for generalized Ellis-Bronnikov solution in the scalar-Einstein-Gauss-Bonnet $4d$ gravitational model, we consider the Einstein tensor and then the stress-energy tensor.
We have for Einstein equations ($c=G=1$)

\begin{eqnarray}
G_{\mu\nu}  = R_{\mu\nu} - \frac{1}{2}g_{\mu\nu}R = 8\pi T_{\mu\nu}. \label{1A_App} 
\end{eqnarray}

Introduce an orthonormal basis:

\begin{equation}
\quad e^{\left(0\right)} =\sqrt{A\left(u\right)}du,  \quad
e^{\left(1\right)} =\frac{du}{\sqrt{A\left(u\right)}} ,  \quad 
e^{\left(2 \right)} =\sqrt{C\left(u\right)}d\theta, \quad  
 e^{\left(3\right)} = \sqrt{C\left(u\right)}  \sin(\theta) d\varphi .   \label{2A2A_App} 
\end{equation}

In this basis, the stress-energy tensor  frame components $T_{\left(a\right)\left(b\right)}$ 
($a,b = 0, 1, 2, 3$) are related to the Einstein tensor ones by:

\begin{eqnarray}
 (T_{\left(a\right) \left(b\right)}) = \frac{1}{8\pi} (G_{\left(a\right)\left(b\right)}) = e^{2\alpha}\frac{\left(\mu^2 + L^2\right)}{\left(u^2 + L^2\right)^2}{\rm diag}\left(-1, -1, 1, 1\right).  \label{2A_App} 
\end{eqnarray}

The energy condution are just inequalities on $T_{\left(a\right)\left(b\right)}$:

\noindent\textbf{1. Weak Energy Condution (WEC)}: $T_{\left(0\right)\left(0\right)} \geq 0$ and  $T_{\left(a\right)\left(a\right)}  + T_{\left(0\right) \left(0\right)}  \geq 0$ for all $a = 1, 2, 3$; \\
\textbf{2. Null Energy Condution (NEC):} $T_{\left(a\right)\left(a\right)}  + T_{\left(0\right)\left(0\right)}  \geq 0$ for all $a = 1, 2, 3$; \\
\textbf{3. Strong Energy Condution (SEC):} $T_{\left(0\right)\left(0\right)} + \sum_{b=1}^3 T_{\left(b\right)\left(b\right)} \geq 0$ and  $T_{\left(a\right) \left(a\right)}  + T_{\left(0\right)\left(0\right)}  \geq 0$ for all $a = 1, 2, 3$; \\
\textbf{4. Dominant Energy Condution (DEC):} $T_{\left(0\right)\left(0 \right)}  \geq |T_{\left(a\right)\left(a\right)}|$ for all $a = 1, 2, 3$. \\

We see that WEC and DEC are violated since $T_{\left(0\right)\left(0\right)} <0$. 
The NEC and SEC are not satisfied due to 
$T_{\left(0\right)\left(0\right)}  + T_{\left(u\right)\left(u\right)} = 2T_{\left(0\right)(\left(0\right)} < 0$. 
Thus, we see that all energy condutions are violated for generalized Ellis-Bronnikov wormhole solution.

\color{black}

 \end{document}